\input harvmac
\input epsf

\def\cross{\!\times\!}
\def\inner{\!\cdot\!}

\def\mod{\hbox{\rm mod}~}

\def\TT{\widetilde T}

\def\Wrp{\widetilde{\hbox{\sl Wr}}}
\def\Lk{\hbox{\sl Lk}}
\def\Wr{\hbox{\sl Wr}}
\def\Tw{\hbox{\sl Tw}}

\def\bold#1{\setbox0=\hbox{$#1$}%
     \kern-.010em\copy0\kern-\wd0
     \kern.025em\copy0\kern-\wd0
     \kern-.020em\raise.0200em\box0 }

\def\grad{\bold{\nabla}}

\def\thetau{\theta_{\scriptscriptstyle U}}

\lref\Kleinert{
See, for instance, H.~Kleinert, {\sl Path Integrals in Quantum Mechanics
Statistics and Polymer Physics}, (World Scientific, Singapore, 1990), Chap.
16.}
\lref\Kas{J.~K\"as, {\sl et al}, Biophys. J. {\bf 70} (1996) 609.}
\lref\Others{A.~Ott, {\sl et al} Phys. Rev. E {\bf 48} (1993) R1642;
A.~Gittes, {\sl et al}, J. Cell Biol. {\bf 120} (1993) 923; H.~Isambert,
{\sl et al}, J. Biol. Chem. {\bf 195} (1995) 11437.}
\lref\Miller{J.D.~Miller, J. Stat. Phys. {\bf 63} (1991) 89.}
\lref\Marko{J.F.~Marko and E.D.~Siggia, Phys. Rev. E {\bf 52} (1995) 2912.}
\lref\deGennes{P.-G.~de Gennes, Phys. Lett. A {\bf 38a} (1972) 339; see also
P.G.~de Gennes, {\sl Scaling Concepts in Polymer Physics}, (Cornell
University Press, Ithaca, NY, 1970), Chapter X.}

\lref\Rudnick{B.~Fain, J.~Rudnick and S.~\"Ostlund, {\sl preprint} (1996)
[cond-mat/9610126];
B.~Fain and J.~Rudnick, {\sl preprint} (1997) [cond-mat/9701040].}
\lref\Julicher{F.~Julicher, Phys. Rev. E {\bf 49} (1994) 2429.}
\lref\Fuller{F.B.~Fuller, Proc. Nat. Acad. Sci. USA {\bf 68} (1971) 815.}
\lref\MH{N.D.~Mermin and T.L.~Ho, Phys.
Rev. Lett. {\bf 36} (1976) 594.}
\lref\LCBO{R.D.~Kamien, J. Phys. II France {\bf 6} (1996)
461 [cond-mat/9507023].}
\lref\Fullerii{F.B.~Fuller, Proc. Natl. Acad. Sci. USA {\bf 75} (1978)
3557.}
\lref\MP{
R.S.~Millman and G.D.~Parker, {\sl Elements of Differential Geometry},
(Prentice-Hall, Englewood Cliffs, NJ, 1977).}
\nfig\fone{Summary of geometry and notation used.  The curve of interest
is $\Gamma$ with capping surface $M$.  The unit tangent vector $\bf T$ to
the curve can be mapped to the surface of the unit sphere.  The area
swept out by the tangent vector on the tangent spherical map is $A$, while
$\bf t$ is the tangent vector to the curve (the tangent indicatrix) lying
on the surface of the sphere.}

\Title{}{Local Writhing Dynamics}

\centerline{
Randall D. Kamien\footnote{$^\ddagger$}
{email: {\tt kamien@lubensky.physics.upenn.edu}}}
\smallskip\centerline{\sl Department of Physics and
Astronomy, University of Pennsylvania, Philadelphia,
PA 19104}
We present an alternative local definition of the writhe of a self-avoiding
closed loop which differs from the traditional non-local definition by
an integer.  When studying dynamics this
difference is immaterial.  We employ a formula due to Aldinger, Klapper
and Tabor for the change in writhe and propose a set of local, link preserving
dynamics in an attempt to unravel some puzzles
about actin.

\Date{13 March 1997; modified 4 April 1997}
Fuller's
ubiquitously used relation between the link, twist, and writhe of a closed
ribbon \Fuller\
has played a role in a great deal of work on the equilibrium statistical
mechanics of DNA \refs{\Marko,\Julicher,\Rudnick}.  While the
traditional definition of writhe unambiguously
depends on the shape of the DNA backbone, it suffers from being non-local
-- it depends on the {\sl entire} conformation.
This non-locality is merely a technical complication for equilibrium physics
but is a severe limitation for the study of dynamics.  In this letter we
present
an alternative, local expression for writhe which differs from the usual
expression for writhe only by an integer.  Since continuous dynamical evolution
cannot smoothly change an integer, from the point of view of dynamics this
new local expression is adequate.  We will, in addition, propose a local,
dynamical link conservation law for a closed curve.
Though we will
not derive this conservation law from first principles dynamics, using
arguments
based only on the locality
of interactions along a twist-storing polymer such as DNA, we will
argue that at some scale it is appropriate.

The classic
result relates the linking number $\Lk$ of the
two sugar-phosphate backbones of DNA to two quantities, the twist $\Tw$,
which is a measure of the rate at which one backbone twists around the
other, and the writhe $\Wr$, which is a measure of how twisted in space the
average backbone is.  These three quantities are related by
\eqn\thething{\Lk = \Tw + \Wr.}
The double helix may be described by the average backbone curve
${\bf R}(s)$ with unit tangent vector
${\bf T}(s)$ and by a unit vector ${\bf U}(s)$ perpendicular to ${\bf T}(s)$
which points from the ${\bf R}(s)$ to one of the two backbones.  In this case
twist and writhe are
\eqn\etwis{\Tw\equiv \oint ds\, {\bf T}(s)\times{\bf U}(s)\inner\dot{\bf U}(s)}
and
\eqn\ewritheold{\Wr \equiv {1\over 4\pi}
\oint ds\,\oint ds'\,{\left[{\bf R}(s) - {\bf R}(s')\right]\inner\dot{\bf
R}(s)\cross
\dot{\bf R}(s')\over \left\vert {\bf R}(s)-{\bf R}(s')\right\vert^3},}
where $\dot X(s)\equiv {dX\over ds}$ \Kleinert\ and $s$ measures the
arc-length along the curve.
We will first suggest a different {\sl local}
expression for writhe $\Wrp$ which we will then
show differs from the usual definition of writhe $\Wr$ by an integer.
While there is no new mathematics in this derivation it is useful to have
a local expression which captures much of the information of writhe.  A similar
resulting expression was first used in the context of
Fermi-Bose transmutation by Polyakov \ref\POL{A.M.~Polyakov, Mod. Phys. Lett.
{\bf 3} (1988) 325.}.

We consider a closed curve $\Gamma$ parameterized by ${\bf R}(s)$ with unit
tangent vector ${\bf T}(s)$ and choose
${\bf e}_1(s)$ and ${\bf e}_2(s)$ along the curve to form an orthonormal triad
with ${\bf T}(s)$:
$\left\{{\bf e}_1(s),{\bf e}_2(s),{\bf T}(s)\right\}$.
Parameterizing the ``difference direction''
${\bf U}(s)$ in terms of ${\bf e}_1$ and ${\bf e}_2$
we have
\eqn\eU{{\bf U}(s)=\cos\thetau(s)\;{\bf e}_1(s) +\sin\thetau(s)\;{\bf e}_2(s).}
The total twist as defined in \etwis\ is the integral of a local ``twist
density'' which we can compute in terms of $\thetau$ and ${\bf e}_i$:
\eqn\etotaltw{\Tw=\int ds\,{{\cal T}\!w} \equiv \oint ds\,{\bf T}\cross{\bf
U}\inner\dot{\bf U}=
 \oint ds\, \left\{\partial_s\thetau
- e_1^\alpha\partial_s e_2^\alpha\right\}}
The first term on the right hand side of \etotaltw\ is necessarily
$2\pi$ times an integer.  What about the second term?  If we
imagine that ${\bf e}_i(s) = {\bf e}_i\big({\bf R}(s)\big)$ are defined
everywhere in space then we define:
\eqn\eintbyparts{
2\pi\Wrp\equiv\oint ds\,e_1^\alpha\partial_s e_2^\alpha = \oint_\Gamma dR_i\,
e_1^\alpha\partial_i
e_2^\alpha.}
Since ${\bf e}_i$ are only defined on the curve, derivatives
in space $\partial_i {\bf e}_2$
are not necessarily well-defined.
However, since we are only interested in this
derivative along the curve, we can make sense of \eintbyparts .
This expression appears to be
a {\sl local} expression for something which we would like, considering
\etotaltw , to interpret as writhe (times $2\pi$).  At first glance adding
the orthonormal triad appears similar to the ``generalized Frenet-Serret''
frame introduced by Goriely and Tabor \ref\GT{A.~Goriely and M.~Tabor, Phys.
Rev. Lett. {\bf 77} (1996) 3557;
Physica D, {\sl to appear} (1997).}.  In the following, we will
propose a dynamics that makes {\sl no reference} to the additional
vectors ${\bf e}_1$ and ${\bf e}_2$.

Suppose the curve $\Gamma$ is the
boundary of a surface $M$ (we can find such a surface for an unknotted
closed curve).
Since ${\bf T}\in S^2$ and $\pi_1(S^2)=0$ we can find a continuous vector field
${\bf\TT}$
on $M$ which is equal to unit tangent vector ${\bf T}$ on the boundary $\Gamma$
\ref\Ed{We assume that we can find a {\sl differentiable}
function ${\bf\TT}$ and do not consider any pathological cases.
Similar results hold
in higher dimensions.  See, for instance, E.~Witten, Commun. Math. Phys. {\bf
92} (1984) 455.}.  In figure 1 we summarize the geometry and notation.
Using Stoke's theorem we then have
\eqn\estoke{\oint_\Gamma dR_i\, e_1^\alpha\partial_ie_2^\alpha =
\int_M dS_i\,\left[\grad\times e_1^\alpha\grad e_2^\alpha\right]_i.}
The integrand appearing in the integral over $M$ has significance in both two
and three dimensions.  If
${\bf e}_1$ and ${\bf e}_2$ are local
tangent vectors to a two-dimensional surface embedded in three dimensions, then
the curl of the ``connection'' $\bold{\omega}\equiv e_1^\alpha\grad
e_2^\alpha$ is the Gaussian curvature of the surface \ref\NELii{D.R.~Nelson and
L.~Peliti,  J. Phys. (Paris) {\bf 48} (1987) 1085.}
while in three dimensions
if ${\bf e}_1$ and ${\bf e}_2$ are perpendicular to, for instance, the nuclear
spin in $^3$He or the long nematic director in a biaxial nematic then the curl
of $\bold{\omega}$ is the geometric realization of a topological fact: in
a biaxial nematic $+1$ disclinations {\sl cannot} escape into the third
dimension
as in uniaxial nematics \LCBO .  The curl of the $\bold{\omega}$ is
\eqn\ecurlthing{\epsilon_{ijk}\partial_j\omega_k=
\left[\grad\times e_1^\alpha\grad e_2^\alpha\right]_i
={1\over 2}\epsilon_{ijk}\epsilon_{\alpha\beta\gamma}
\TT^\alpha\partial_j\TT^\beta
\partial_k\TT^\gamma,}
the famed Mermin-Ho relation \MH .  In the following we will show that
there is a relation between
our proposed expression for writhe and the more traditional expression
\ewritheold .

Nearly twenty years ago \Fullerii\ Fuller showed
that the writhe $\Wr$ of a curve differed by an integer from the signed
area swept out by the unit tangent vector ${\bf T}$ on the unit sphere.  For
completeness we review that result here.  As before,
let ${\bf T}(s)$ be the unit
tangent vector to the curve ${\bf R}(s)$ and ${\bf U}(s)$ be
the vector pointing along the ribbon with ${\bf U}(s)\inner{\bf T}(s)=0$.
${\bf T}(s)$ traces out
a curve on the unit sphere (the so-called ``tangent indicatrix'').
Let the tangent to this curve be ${\bf t}(s)$.  Since ${\bf U(s)}$ is
perpendicular to ${\bf T}(s)$, it lies in the tangent plane of the
unit sphere at ${\bf T}(s)$.  Applying the Gauss-Bonnet theorem \MP\
to the region $A$ enclosed
by the curve on the unit sphere gives
\eqn\egb{\int_A d^2\!\sigma\,K + \int_{\partial A}ds\,\kappa_g = 2\pi}
where $K$ is the Gaussian curvature and $\kappa_g$ is the geodesic curvature,
defined as the
rate of change of ${\bf t}(s)$ in the tangent plane of the sphere.
Locally parameterizing $A$ by the orthonormal vectors ${\bf e}_1(\sigma_1,
\sigma_2)$ and
${\bf e}_2(\sigma_1,\sigma_2)$
(with ${\bf e}_1(\vec\sigma)\cross{\bf e}_2(\vec\sigma) =
{\bf n}(\vec\sigma_1)$, where ${\bf n}$ is the surface normal),
we may write the unit vectors ${\bf t}$ and ${\bf U}$ as
\eqn\etandU{\eqalign{
{\bf t}(s) &=\cos\theta_t(s)\;{\bf e}_1(s) +\sin\theta_t(s)\;{\bf e}_2(s)\cr
{\bf U}(s) &=\cos\thetau(s)\;{\bf e}_1(s) +\sin\thetau(s)\;{\bf e}_2(s)\cr
}}
where ${\bf e}_i(s) = {\bf e}_i(\tilde\sigma(s))$ and $\tilde\sigma(s)$
is the parameterization of the boundary in terms of the surface
co\"ordinates $\vec\sigma$.  Since on $\partial A$ the surface
normal is ${\bf T}(s)$, the geodesic curvature is
$\kappa_g={\bf T}\cross{\bf t}\inner\dot{\bf t}$
while the twist density ${\cal T}\!w$ is
defined as $2\pi {{\cal T}\!w}={\bf T}\cross{\bf U}\inner\dot{\bf U}$.
Using the orthonormality of $\{{\bf e}_1,{\bf e}_2,{\bf T}\}$ we have
\eqn\ethecurvature{\kappa_g = \partial_s\theta_t(s) - e_1^\alpha\partial_s
e_2^\alpha,}
and
\eqn\ethetwist{2\pi {{\cal T}\!w} = \partial_s\thetau(s) - e_1^\alpha\partial_s
e_2^\alpha.}
Since the difference is $2\pi {{\cal T}\!w}
- \kappa_g = \partial_s(\thetau-\theta_t)$
after one circuit about the curve the integral of the difference
changes by an integer multiple of $2\pi$.  Using \egb\ and the fact
that the Gaussian curvature of the unit sphere is $K=1$, Fuller found that
\eqn\efuller{{A\over 2\pi} + \Tw \equiv 0~\mod 1,}
where $A$ is the signed area of the region enclosed by ${\bf T}(s)$ on the
unit sphere and $\Tw \equiv \int ds\,{\cal T}\!w$.  Comparing this to
$\Lk=\Tw+\Wr$,
we see that
\eqn\ewri{\Wr\equiv {A\over 2\pi}~\mod 1,}
which establishes Fuller's result.  Upon using Stoke's theorem and the
Mermin-Ho relation we have:
\eqn\ewrrr{2\pi\Wrp = \oint ds\, e_1^\alpha\partial_se_2^\alpha =
{1\over 2}\int_M
dS_i\,\epsilon_{ijk}\epsilon_{\alpha\beta\gamma}\TT^\alpha\partial_j
\TT^\beta\partial_k\TT^\gamma}
which is {\sl precisely} the area swept out by the curve ${\bf T}$ on the
unit sphere!  Thus we see that
\eqn\emodd{\Wr \equiv \Wrp~\mod 1.}
While the total amount of writhe including the integral part is important
for calculating ground states of a particular twisted ribbon
\refs{\Marko,\Rudnick,
\Julicher}\ when considering changes in writhe the integer is less important.
In
particular to study dynamics a local ``writhe density'' is probably essential.

To this end, a useful
result \ref\knot{J.~Aldinger, I.~Klapper and M.~Tabor, J. Knot Theory
Ramifications {\bf 4} (1995) 343.}\ for the change in writhe of a curve
as a function of some deformation $\lambda$
\eqn\euseful{{\partial_\lambda \Wr(\lambda)} = {1\over 2\pi}
\oint ds\, {\bf T}(s,\lambda)
\inner\left[\partial_\lambda{\bf T}(s,\lambda)\cross\partial_s{\bf T}(s,
\lambda)\right],}
can be used.  Note that this result follows from the preceding discussion:
\euseful\ (multiplied by $d\lambda$) is the differential in the area swept out
by the tangents of
two closed curves with tangent vectors
${\bf T}(s,0)$ and ${\bf T}(s,d\lambda)$.  Since the twist is really
the local torsional strain of the polymer, we denote it as $\Omega(s,t)$ and
then we have
\eqn\ederiv{\partial_t \Lk = \oint ds\,\left\{\partial_t\Omega
+ \partial_t{\bf T}\inner\left[\partial_s{\bf T}\cross{\bf T}\right]\right\}+
\partial_t n}
where $n$ is the integer difference between $\Wr$ and $\Wrp$.  Since continuous
evolution
cannot lead to discontinuous changes in the integer $n$ and since link is
conserved we have
\eqn\edrri{0=\oint ds\,\left\{\partial_t\Omega
+ \partial_t{\bf T}\inner\left[\partial_s{\bf T}\cross{\bf T}\right]\right\}.}
This conservation law need not be satisfied locally: the curve can twist in one
place and writhe at some distant location to satisfy \edrri .  While
mathematically
consistent this is not a physically plausible effect.  We expect that the
linking
number is conserved {\sl locally} and only changes through the diffusion of
some
``link current'' $j$.  With this assumption we have the local conservation law
\eqn\elinck{\partial_s j = \partial_t\Omega + \partial_t{\bf
T}\inner\left[\partial_s{\bf T}
\cross{\bf T}\right],}
which satisfies \edrri .  Note that although we introduced the
framing vectors ${\bf e}_i$ they do not appear in the conservation
law.  The invariance of $\Wrp$ with respect to the choice of ${\bf e}_i$
only holds for a closed curve.  Nonetheless the form of \elinck\
suggests that for dynamics we may apply the conservation law to
{\sl open curves} .

We consider now the relaxational Rouse dynamics of
a closed, twist-storing polymer with a free energy
\eqn\efree{F={1\over 2}\int ds\,\left\{A\left(\partial_s{\bf T}\right)^2 +
C\Omega^2
-\lambda\left({\bf T}^2-1\right)\right\}}
where $A$ and $C$ are the bend and twist elastic constants, respectively,
and $\lambda$ is the Lagrange multiplier which
enforces the constraint that $\bf T$ be a unit vector \ref\LG{
R.E.~Goldstein and S.A.~Langer, Phys. Rev. Lett. {\bf 75} 1094 (1995).}.
The dynamical equations for $\bf T$ and $\Omega$ are
\eqna\dynas{
$$\eqalignno{
\partial_t\Omega & = -{\Gamma_{\!\scriptscriptstyle\Omega}}{\delta F\over\delta
\Omega} =
-C\Gamma_{\!\scriptscriptstyle\Omega}\Omega&\dynas a\cr
\partial_t{\bf T} & = -{\Gamma_{\!\scriptscriptstyle T}}{\delta F\over\delta
{\bf T}}
= A\Gamma_{\!\scriptscriptstyle T}\partial^2_s{\bf T} +
\Gamma_{\!\scriptscriptstyle T}\lambda{\bf T}&\dynas b\cr
0&={\bf T}^2 -1,&\dynas c\cr}$$}
where $\Gamma_{\!\scriptscriptstyle\Omega}$ and $\Gamma_{\!\scriptscriptstyle
T}$ are dissipation constants.
Using \dynas{}\ we may rewrite
\elinck\ as
\eqn\elincktwo{\partial_s j = -C\Gamma_{\!\scriptscriptstyle\Omega}\Omega -
A\Gamma_{\!\scriptscriptstyle T}{\bf T}
\inner\left[\partial_s{\bf T}\cross\partial_s^2{\bf T}\right].}
We recognize the second term on the right-hand side of \elincktwo\ as
$\kappa^2(s)\tau(s)$ where $\kappa(s)$ is the curvature of the
curve and $\tau(s)$ is the torsion
\nref\KT{I.~Klapper and M.~Tabor, J. Phys. A {\bf 27}, 4919 (1994).}.
Note that while usually the
torsion is ill-defined for a curve with zero curvature, there is no
problem here because of the explicit factor of $\kappa^2$.  This formulation
differs from the work in \KT\ by the introduction of a link density
current which allows a local constraint.
Thus
we see that the ``link current'' obeys
\eqn\elicu{\partial_s j = -C\Gamma_{\!\scriptscriptstyle\Omega} \Omega -
A\Gamma_{\!\scriptscriptstyle T} \kappa^2\tau.}
This form of the link current allows us to locally understand
how link moves: link moves when there is a twist and when the curve
is non-planar (i.e. $\tau\ne 0$).
Assuming a Fick's law form for the current $j=-D\partial_s{\cal L}k$
in terms of the
link density ${\cal L}k$, we finally have
\eqn\efinal{D\partial^2_s{\cal L}k = C\Gamma_{\!\scriptscriptstyle\Omega}
\Omega +
A\Gamma_{\!\scriptscriptstyle T} \kappa^2\tau.}
Note that Fick's law implies that if the link is uniformly spread along
the polymer then the current is constant.

As a simple application of \elinck\ we can consider a closed ribbon
of radius
$r=\kappa^{-1}$ lying in the
$xy$-plane.
If we consider relaxational dynamics around the curved ground state
we see that in this case \efinal\ will depend {\sl linearly} on
${\bf T}$ through the torsion $\tau$ (as opposed to quadratically
if $\kappa$ were $0$).
We can consider the dynamics of the ribbon in extreme cases.
The diffusion constant $D$ can be large (infinite) or small (zero).
Additionally the twist modulus $C$ can be large (infinite) or small (zero).
If the diffusion constant is infinite then
\elinck\ is not much of a constraint:  any writhe or twist deformation
can be quickly compensated for by small local link deformations.
This is the limit in which
dynamics is non-local and in which the only constraint is \edrri .

The case
of zero diffusion is more interesting and corresponds to ``ultra-local''
dynamics in which link is conserved point by point.  In this case
we have $j=0$ and so
\eqn\eslow{C\Gamma_{\!\scriptscriptstyle\Omega}\Omega =
-A\Gamma_{\!\scriptscriptstyle
T}
\kappa^2\tau.}
If the twist modulus is $0$ or small then it is easy to change
$\Omega$ and this
is hardly a constraint.  However, if the twist modulus is large or
infinite then
\eslow\ implies that if the curve starts planar then {\sl it remains planar}
since $\tau=0$ and $\kappa\ne 0$.
This is, of course, only true if the polymer has an everywhere
non-zero curvature.  We note that this observation may explain some puzzling
experiments
on actin, another double-helical polymer \refs{\Kas,\Others}.  Though
$\Lk=\Tw+\Wr$ applies
only to closed curves, we would expect the local dynamics of closed
twist-storing
polymers to be the same as that for similar open polymers.
In those experiments
on open actin strands there
were two surprising observations.  The first was that in \Kas ,
at long wavelengths,
the persistence length seems to grow from
around $2 \mu m$ to roughly $10 \mu m$ though
it remained uniform at $2 \mu m$ until some crossover wavenumber $q_c$.
The
experiment was done by observing actin filaments oscillating in a
sample volume.  In
order to distinguish in-plane motion from out of plane-motion, an actin
filament
which did not completely remain in focus for the duration of the sampling was
not used in the data analysis.  It is rather surprising that a polymer
would remain in a single plane for an extended period
\ref\pc{The cover slides in \Kas\
were considered too far apart (up to $10 \mu m$) to have imparted a significant
effect on the actin.}.
In the light of \eslow\ a possible explanation emerges: if an
actin filament has
zero average curvature, then, on average, \elinck\ is not
a linear constraint since $\kappa=0$.  However if the longest wavelength
undulation
modes have not equilibrated then it is possible for $\langle\kappa\rangle\ne
0$.  This
could explain {\sl both} effects.  The long wavelength modes do not have good
statistics which is why they appear to have an anomalous persistence length
and, at
the same time, the non-equilibration of these modes forces the torsionally
stiff actin filament
to remain in the plane for some time until link can diffuse in from the ends.

As a mathematical aside for the {\sl cogniscenti} we note that when continuing
$\bf T$
into the region $M$ there is an ambiguity since $\pi_2(S^2)={\cal Z}$.
However, the integrand in \ewrrr\ is the skyrmion density
\ref\TR{H.-R.~Trebin, Adv. Phys. {\bf 31} (1982) 195.}, which,
when integrated over the entire region can differ by $2\pi$ times an integer
for interior continuations which are homotopically distinct.
This is not a problem since there is already a $\mod 1$ ambiguity
between $\Wr$ and $A/(2\pi)$.  Finally, we might consider the
case of a knotted curve.  In this case the curve can be unknotted
by adding to the curve small loops.  It can be shown in this case
that the addition of a small segment of curve will change
$\Wrp$ by an integer, which is therefore unimportant for dynamics.

I would like to thank J.~K\"as, T.C.~Lubensky, J.~Marko,
P.~Nelson and D.~Pettey for discussions.  I especially thank T.R.~Powers
for many conversations and a critical reading of this manuscript.
This work was supported through NSF Grant DMR94-23114.

\listrefs
\listfigs
\nopagenumbers
\epsfysize=8.8truein
\centerline{\epsfbox{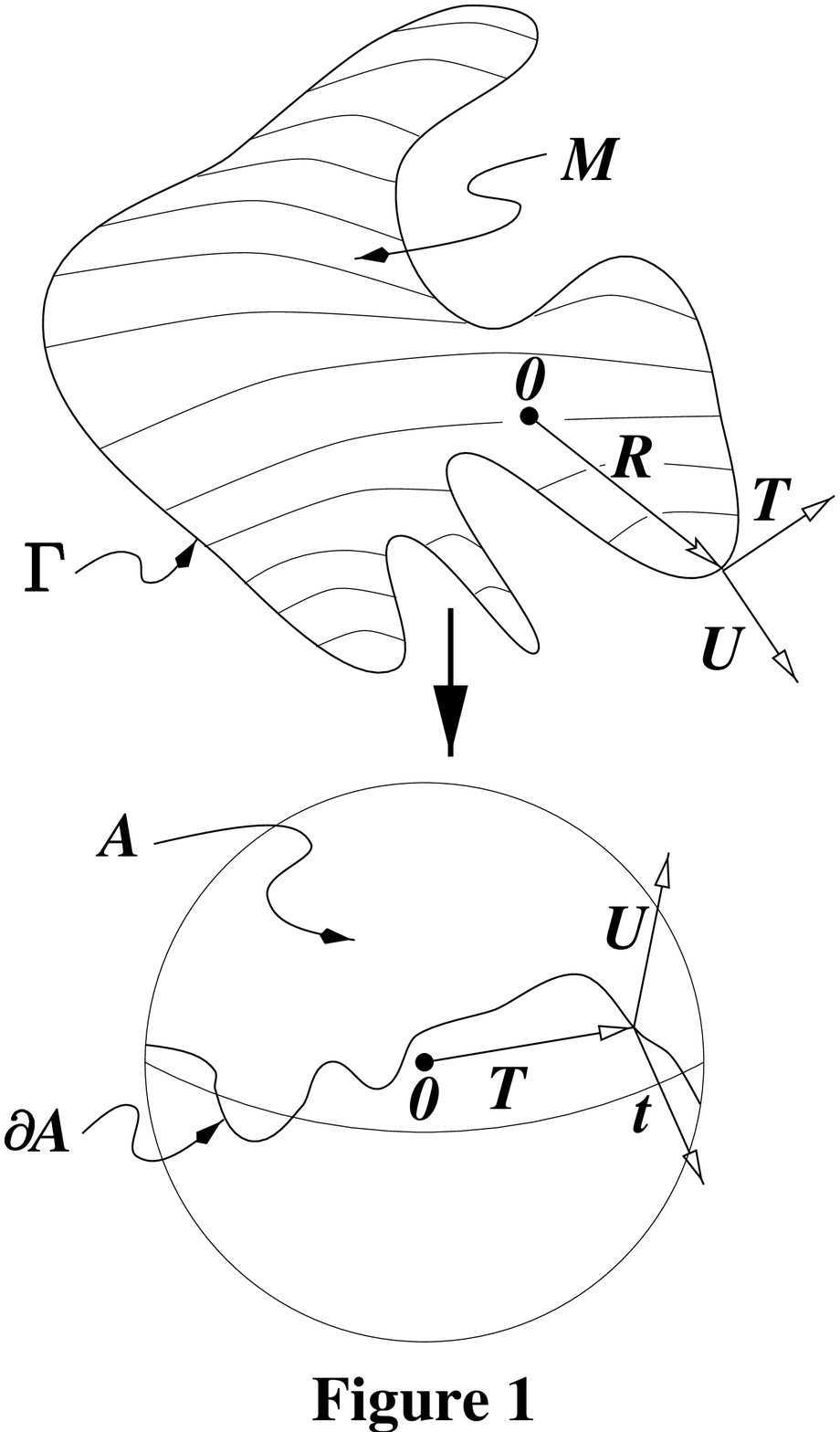}}

\bye